\documentclass[10pt]{iopart}
\usepackage{graphicx}
\usepackage[figuresright]{rotating}

\begin{document}

\title[Power corrections to the Bjorken sum rule]{Modeling power corrections
to the Bjorken sum rule for the neutrino structure function $F_1$}

\author{C. Weiss}

\address{Institut f\"ur Theoretische Physik, 
Universit\"at Regensburg, \\
D--93053 Regensburg, Germany \\
(christian.weiss@physik.uni-regensburg.de)}

\begin{abstract}
Direct measurements of the the structure functions $F_1^{\nu p}$ 
and $F_1^{\nu n}$ at a neutrino factory would allow for an accurate
extraction of $\alpha_s$ from the $Q^2$--dependence of the Bjorken 
sum rule, complementing that based on the Gross--Llewellyn-Smith 
sum rule for $F_3$. We estimate the power ($1/Q^2$--) corrections to 
the Bjorken sum rule in the instanton vacuum model. For the reduced matrix 
element of the flavor--nonsinglet twist--4 operator 
$\bar u \, g \widetilde G_{\mu\nu} \gamma_\nu \gamma_5 u - (u \rightarrow d)$ 
we obtain a value of $0.18 \, {\rm GeV}^2$, in good agreement with the 
QCD sum rule calculations of Braun and Kolesnichenko. Our result allows 
to reduce the theoretical error in the determination of $\alpha_s$.
\end{abstract}




The precise determination of the strong coupling $\alpha_s$ remains a
prime objective of particle physics. One way to measure $\alpha_s$ 
is through the $Q^2$--dependence of the Gross--Llewellyn-Smith (GLS) 
sum rule for the isoscalar neutrino structure function 
$F_3^{\nu p} + F_3^{\nu n}$ \cite{Gross:1969jf}. With the perturbative 
corrections known exactly up to order $\alpha_s^3$ \cite{Gorishnii:1985xm}, 
scheme and scale ambiguities can be minimized \cite{Chyla:1992cg}, 
and $\alpha_s$ was extracted 
from QCD fits to data combined from various experiments (CCFR, CERN, 
IHEP) \cite{Kim:1998ki}. 
A closely related sum rule is the Bjorken sum rule for the isovector 
structure function $F_1$
\cite{Bjorken:1967px}
\begin{equation}
\int_0^1 dx \; \left[ F_1^{\nu n} (x, Q^2) - F_1^{\nu p} (x, Q^2) \right]
\;\; = \;\; 1 - \frac{2}{3} \frac{\alpha_s (Q^2)}{\pi} + \ldots .
\label{Bj_phenom}
\end{equation}
The perturbative corrections have also been computed up 
to order $\alpha_s^3$ \cite{Gorishnii:1983gs} (the $\alpha_s^4$ 
contribution was estimated in Refs.\cite{Kataev:1995vh}).
This sum rule has so far not been tested experimentally. 
While $F_1$ is theoretically related to $F_2$ 
by the Callan--Gross relation, only recently have experiments begun
to extract $F_1^{\nu} (x, Q^2)$ directly from the cross section.
Measurements have been reported by the CHORUS experiment at CERN 
\cite{Oldeman:pe} and the CCFR--NuTeV Collaboration at 
Fermilab \cite{Yang:2000uc}; however, the results cover only a limited range 
in $x$. High--statistics experiments at a neutrino factory would allow
to separate the various components of the cross section, 
including $F_1 (x, Q^2)$. This would offer the
possibility of using the Bjorken sum rule (\ref{Bj_phenom}) for an 
independent accurate extraction of $\alpha_s$, 
complementing that from the GLS sum rule \cite{Mangano:2001mj}.

An important issue in the determination of $\alpha_s$ from QCD fits
to both the GLS and the Bjorken sum rule are power ($1/Q^2$--)
corrections. Ideally, one would
determine the size of these corrections phenomenologically, from
the fit to the data. However, correlations of $\alpha_s$ with the 
coefficient of the $1/Q^2$ correction increase with the order of the 
perturbative expansion, as a result of which the accuracy of the extracted 
$\alpha_s$ does not improve with increasing order \cite{Chyla:1992cg}. 
A more promising approach is to rely on ``advance knowledge'' of the size 
of the power correction from theoretical estimates.
Aside from the known target mass corrections the $1/Q^2$ corrections 
to the GLS and Bjorken sum rules are due to non-perturbative quark--gluon 
correlations in the nucleon. The coefficients of the $1/Q^2$--corrections
are given by 
$-(8/9) \langle\langle O_{\rm S (NS)} \rangle\rangle$, respectively, where 
$\langle\langle O \rangle\rangle$ is the reduced proton matrix elements of 
the twist--4 operator \cite{Shuryak:1982kj}
\begin{equation}
O_{\alpha} \;\; = \;\; \bar q \; 
g \widetilde G_{\alpha\beta} 
\; \gamma_\beta \gamma_5 \; q ,
\hspace{3em}
\langle p | O_\alpha | p \rangle \;\; = \;\;
2 p_\alpha \; \langle\langle O \rangle\rangle ,
\label{O}
\end{equation}
and S and NS denote the flavor--singlet and nonsinglet combinations
$\bar u \ldots u \pm \bar d \ldots d$. Here $\widetilde G_{\alpha\beta} \equiv 
\epsilon_{\alpha\beta\gamma\delta} G_{\gamma\delta} / 2$ is the dual 
gluon field strength tensor.
%
%
\begin{figure}[t]
\includegraphics[width=12.33cm,height=1.8cm]{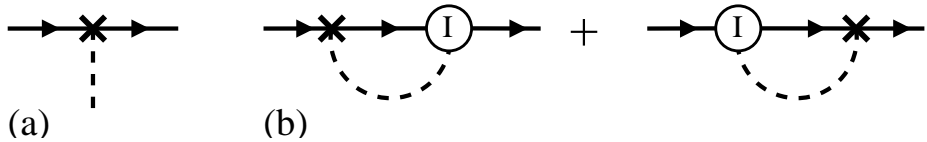}
\caption[]{(a) The twist--4 quark--gluon operator, Eq.(\ref{O}).
(b) Quark--gluon correlations induced by single instantons in 
the instanton vacuum model. The dashed line denotes the gauge field 
of the instanton, the circle the chirality--flipping fermion vertex 
induced by the zero mode of the instanton. (Shown is the situation for 
a single light quark flavor ($N_f = 1$); in general the instanton
vertex is a $2 N_f$--point fermion vertex.)}
\label{fig_effop}
\end{figure}

The matrix elements (\ref{O}) were estimated by Braun and Kolesnichenko 
using QCD sum rules \cite{Braun:1987ty}, see Table~\ref{tab_1}. 
An alternative approach is based on the picture of the QCD vacuum
as a ``medium'' of instantons ---
topological fluctuations of the gauge fields. This picture explains 
the dynamical breaking of chiral symmetry in QCD and a host of 
phenomenological data on hadronic correlation 
functions \cite{Shuryak:1982ff,Diakonov:1984hh}, 
and is supported by lattice simulations. Our estimate
of higher--twist matrix elements is based on the analytic approach 
to the instanton vacuum by Diakonov and Petrov 
\cite{Diakonov:1984hh}. The average size of the instantons 
in the vacuum is $\bar\rho \approx 0.3 \, {\rm fm}$, 
while their average distance is $\bar R \approx 1 \, {\rm fm}$.
The fact that the instanton medium is dilute, 
$\bar\rho / \bar R \approx 1/3 \ll 1$, is of crucial importance for this
picture. It allows for a systematic classification of non-perturbative 
effects generated by instantons. In leading order of $\bar\rho / \bar R$
quark--gluon correlations as measured by the operator (\ref{O}) are 
induced by {\it single} instantons, see Fig.~\ref{fig_effop}. 
By coupling to the instanton the quark--gluon QCD operator 
(\ref{O}) (normalized at the scale $\mu \sim \bar\rho^{-1} = 600\, {\rm MeV}$)
turns into a chirality--flipping ``effective quark operator'', whose 
matrix element is to be evaluated in the low--energy effective theory 
derived from the instanton vacuum, characterized by a dynamical mass of the 
quarks and the appearance of Goldstone boson modes, the pions
(for details see Refs.\cite{Diakonov:1996qy,Balla:1998hf}). This
effective theory has extensively been tested and shown to reproduce
the ``chiral phenomenology'' of strong interactions at low energies.
In particular, it describes the nucleon as a ``soliton'' of the pion field 
in the large--$N_c$ limit \cite{Diakonov:1988ty}.

The flavor--singlet nucleon matrix element was estimated in the
instanton vacuum model in Ref.~\cite{Balla:1998hf}, see Table~\ref{tab_1}. 
Here we report about an estimate of the flavor--nonsinglet matrix element, 
which determines the $1/Q^2$ corrections to the Bjorken sum rule. 
The ``effective quark operator'' obtained from the 
single--instanton contribution of Fig.~\ref{fig_effop} has the quantum 
numbers of the isovector vector current of the effective low--energy theory,
and the twist--4 nucleon matrix element is proportional to the 
nucleon vector charge, with a coefficient of the order of the square 
of the inverse instanton size, $\bar\rho^{-2}$ 
(details will be given elsewhere). We find a 
numerical value of $\langle\langle O_{\rm NS} \rangle\rangle = 
0.5 \, \bar\rho^{-2} = 0.18 \, {\rm GeV}^2$. The instanton vacuum 
results for both the flavor--singlet and nonsinglet matrix elements
are in good agreement with the QCD sum rule predictions of 
Ref.\cite{Braun:1987ty}, which is very encouraging, given the 
general difficulties with modeling higher--twist matrix elements.
The accuracy of the QCD sum rule results was estimated at about $\pm30\% $
\cite{Braun:1987ty}. The theoretical error of the instanton predictions 
is difficult to quantify; we expect it to be not larger than $\pm 50 \% $.

A recent simulation of the $Q^2$--dependence of the Bjorken sum rule
incorporating twist--4 corrections as estimated from QCD sum 
rules \cite{Braun:1987ty} found the theoretical error of the extracted 
$\alpha_s$ to be dominated by the uncertainty of the twist--4
matrix element \cite{Alekhin:2002pj}. Our result indicates that one
can be more confident about the magnitude of this matrix element.
This makes the idea of an accurate measurement of this sum rule
at a neutrino factory even more attractive \cite{Mangano:2001mj}. 
Note also that analyses of power corrections to 
$x (F_3^{\nu N} + F_3^{\bar\nu N})$ within the infrared renormalon model, 
based on the IHEP and CCFR data \cite{Alekhin:2001zj}, suggest a large 
negative twist--4 contribution to the GLS sum rule \cite{Alekhin:private}, 
which would be consistent with the theoretical estimates of 
$\langle\langle O_{\rm S} \rangle\rangle$
quoted in Table~\ref{tab_1}. Finally, the instanton vacuum can be used 
to model also power corrections to polarized electron/muon structure functions
\cite{Balla:1998hf,Lee:2001ug}. First results of a 
comparison of the instanton predictions with the higher--twist contribution 
to $g_1$ extracted from QCD fits are very encouraging \cite{Sidorov:private}.
%
%
\begin{table}
\begin{center}
\begin{tabular}{|l|c|c|}
\hline
& $\langle\langle O_{\rm S} \rangle\rangle / {\rm GeV^2}$  
& $\langle\langle O_{\rm NS} \rangle\rangle / {\rm GeV^2}$  \\
\hline
QCD sum rules \cite{Braun:1987ty} & 0.33 & 0.15 \\
Instanton vacuum \cite{Balla:1998hf} & 0.32 & 0.18 \\
\hline
\end{tabular}
\caption[]{Theoretical estimates of the twist--4 matrix element 
(\ref{O}), determining $1/Q^2$ corrections to the GLS (S) and 
Bjorken (NS) sum rules.}
\label{tab_1}
\end{center}
\end{table} 

I am grateful to S.~I.~Alekhin, A.~L.~Kataev, and A.~V.~Sidorov
for valuable discussions and correspondence. C.W.\ is a Heisenberg 
Fellow (DFG).


\begin{thebibliography}{99}
%
%
\bibitem{Gross:1969jf}
D.~J.~Gross and C.~H.~Llewellyn Smith,
Nucl.\ Phys.\  {\bf B14} (1969) 337.
%
%
\bibitem{Gorishnii:1985xm}
S.~G.~Gorishnii and S.~A.~Larin,
Phys.\ Lett.\ B {\bf 172}, 109 (1986);
S.~A.~Larin and J.~A.~Vermaseren,
Phys.\ Lett.\ B {\bf 259}, 345 (1991).
%
%
\bibitem{Chyla:1992cg}
J.~Chyla and A.~L.~Kataev,
Phys.\ Lett.\ B {\bf 297}, 385 (1992)
[hep-ph/9209213].
%
%
\bibitem{Kim:1998ki}
J.~H.~Kim {\it et al.},
Phys.\ Rev.\ Lett.\  {\bf 81}, 3595 (1998)
[hep-ex/9808015].
%
%
\bibitem{Bjorken:1967px}
J.~D.~Bjorken,
Phys.\ Rev.\  {\bf 163}, 1767 (1967).
%
%
\bibitem{Gorishnii:1983gs}
K.~G.~Chetyrkin, S.~G.~Gorishnii, S.~A.~Larin and F.~V.~Tkachov,
Phys.\ Lett.\ B {\bf 137}, 230 (1984);
S.~A.~Larin, F.~V.~Tkachov and J.~A.~Vermaseren,
Phys.\ Rev.\ Lett.\  {\bf 66}, 862 (1991).
%
%
\bibitem{Kataev:1995vh}
A.~L.~Kataev and V.~V.~Starshenko,
Mod.\ Phys.\ Lett.\ A {\bf 10}, 235 (1995)
[hep-ph/9502348];
M.~A.~Samuel, J.~R.~Ellis and M.~Karliner,
Phys.\ Rev.\ Lett.\  {\bf 74}, 4380 (1995)
[hep-ph/9503411];
D.~J.~Broadhurst and A.~L.~Kataev,
Phys.\ Lett.\ B {\bf 544}, 154 (2002)
[hep-ph/0207261].
%
%
\bibitem{Oldeman:pe}
R.~G.~Oldeman [CHORUS Collaboration],
Nucl.\ Phys.\ Proc.\ Suppl.\  {\bf 79}, 96 (1999).
%
%
\bibitem{Yang:2000uc}
U.~K.~Yang {\it et al.}  [CCFR / NuTeV Collaboration],
hep-ex/0010001.
%
%
\bibitem{Mangano:2001mj}
M.~L.~Mangano {\it et al.}, hep-ph/0105155.
%
%
%
\bibitem{Shuryak:1982kj}
E.~V.~Shuryak and A.~I.~Vainshtein,
Nucl.\ Phys.\  {\bf B199} (1982) 451.
%
%
\bibitem{Braun:1987ty}
V.~M.~Braun and A.~V.~Kolesnichenko,
Nucl.\ Phys.\  {\bf B283} (1987) 723.
%
%
\bibitem{Shuryak:1982ff}
E.~V.~Shuryak, Nucl.\ Phys.\  {\bf B203} (1982) 93;
{\it ibid.}\ {\bf B203} (1982) 116.
%
%
\bibitem{Diakonov:1984hh} 
D.~I.~Diakonov and V.~Y.~Petrov, 
Nucl.\ Phys.\  {\bf B245} (1984) 259;
Nucl.\ Phys.\  {\bf B272} (1986) 457. 
%
%
\bibitem{Diakonov:1996qy}
D.~I.~Diakonov, M.~V.~Polyakov and C.~Weiss,
Nucl.\ Phys.\  {\bf B461} (1996) 539
[hep-ph/9510232].
%
%
\bibitem{Balla:1998hf}
J.~Balla, M.~V.~Polyakov and C.~Weiss,
Nucl.\ Phys.\  {\bf B510} (1998) 327
[hep-ph/9707515].
%
%
\bibitem{Diakonov:1988ty}
D.~I.~Diakonov, V.~Y.~Petrov and P.~V.~Pobylitsa,
Nucl.\ Phys.\  {\bf B306} (1988) 809.
%
%
\bibitem{Alekhin:2002pj}
S.~I.~Alekhin and A.~L.~Kataev, these proceedings [hep-ph/0209165].
%
%
\bibitem{Lee:2001ug}
N.~Y.~Lee, K.~Goeke and C.~Weiss,
Phys.\ Rev.\ D {\bf 65}, 054008 (2002)
[hep-ph/0105173].
%
%
\bibitem{Alekhin:2001zj}
S.~I.~Alekhin {\it et al.}, Phys.\ Lett.\ B {\bf 512}, 25 (2001)
[hep-ex/0104013].
%
%
\bibitem{Alekhin:private} S.~I.~Alekhin, private communication
%
%
\bibitem{Sidorov:private}
A.~V.~Sidorov, private communication
%
%
\end{thebibliography}
\end{document}